\documentclass[twocolumn,12pt,cites,graphicx]{article}
\usepackage{bm}
\usepackage[english]{babel}
\usepackage{amsfonts}
\usepackage{amssymb, amsmath}
\usepackage{epsfig}
\usepackage{amsbsy}
\usepackage{cite}

\title{A unit-cell approach to the nonlinear rheology of\\
biopolymer solutions} 

\author{Pablo Fern\'andez$^\dagger$, Steffen Grosser$^\ast$ 
and Klaus Kroy$^\ast$\\[2ex]
$^{\dagger}$Lehrstuhl f\"ur Zellbiophysik E27, Technische Universit\"at M\"unchen\\
\normalsize James-Franck-Stra\ss{}e 1, D-85748 Garching, Germany\\[1ex]
$^{\ast}$Institut f\"ur Theoretische Physik, Universit\"at Leipzig\\
\normalsize Postfach 100920, D-04009 Leipzig, Germany}

\date{\today}
\begin{document} 

\maketitle 
\renewcommand{\thefootnote}{\fnsymbol{footnote}}

\noindent We propose a nonlinear extension of the standard tube model
  for semidilute solutions of freely-sliding semiflexible polymers.  Non-affine filament
  deformations at the entanglement scale, the renormalisation of
  direct interactions by thermal fluctuations, and the geometry of
  large deformations are systematically taken into account.  The
  stiffening response predicted for athermal solutions of
  stiff rods \cite{doikuzuu} is found to be thermally suppressed. Instead, we
  obtain a broad linear response regime,
    supporting the interpretation of shear stiffening at finite
  frequencies in polymerised actin  solutions  \cite{christine,nonlinear_semmrich08}
  as indicative of  coupling to longitudinal modes. 
  We observe a destabilizing effect of large strains ($\sim 100\%$),
  suggesting shear banding as a plausible explanation for
  the widely observed catastrophic collapse of \emph{in-vitro}
  biopolymer solutions, usually attributed to network damage.
  In combination with
  friction-type interactions, our analysis provides an analytically
  tractable framework to address the nonlinear viscoplasticity of
  biological tissue on a molecular basis.

\section{Introduction}\label{intro}
The remarkable mechanical properties of animal cells and tissues are
attributed to a dense meshwork of semiflexible biopolymers known as
the cytoskeleton. \emph{In-vitro} polymerized biopolymer solutions
provide a relatively well-controlled starting point for a systematic
study of its basic physical properties
\cite{bottomup_cellmech}. Particularly relevant are solutions of
filamentous actin (F-actin), which is the universal cellular
scaffolding element and moreover contributes to cytoskeletal
remodelling and force generation by its treadmilling
polymerisation \cite{braybook}. While conceptionally simpler than crosslinked
networks, pure solutions are commonly regarded as physiologically less
relevant.  However, one may wonder about the rational basis for
numerous recent ``explanations'' of the linear and nonlinear mechanics
of cells and crosslinked \emph{in-vitro} networks as long as we do not
have a sound understanding of these properties for purely entangled
solutions. Arguably, the actin binding proteins present in the
cytoskeleton of living cells mostly cause transient rather than
irreversible crosslinking, the whole network architecture being
constantly fluidized by treadmilling and molecular motor
activity. Indeed, numerous recent studies have established that the
intracellular polymer network in living cells resembles a ``soft
glassy'' \cite{glassy1,cellmech_trepat07}, viscoplastic \cite{monolayer,fernandez2} material
rather than a fixed elastic carcass. Its rheology exhibits
intriguing similarities to that of pure actin solutions thermostated
at low temperatures,
which --surprisingly-- show shear stiffening responses 
despite the absence of crosslinkers  \cite{christine}. 
Accordingly, the long-time
integrity and mechanical stability of the cytoskeleton
seems to call for an explanation in
terms of a merely physically entangled polymer solution, in which
transient crosslinkers and other friction-like interactions affect the
mechanics chiefly through a slowdown of the relaxation. The aim of
this contribution is to lay the foundations for an analytical
implementation of an appropriate minimal model.

The standard approach to the quasi-static mechanics of entangled
semiflexible polymers with hard core interactions employs the
so-called \emph{tube model of semiflexible polymers} --- not to be
confused with the tube model for flexible polymers, which is a
dedicated scheme to deal with dynamics. Its essence is a simple
scaling theory constructed in direct analogy with the popular ``blob
model'' for semidilute flexible polymer solutions \cite{review_kroy}. In
the semiflexible case, anisotropic blobs of length $L_e$ and a
transverse cross-section on the order of $d^2\simeq L^3_e/\ell_p$
account for the anisotropic fluctuations of a weakly bending rod of
persistence length $\ell_p$ \cite{odijk}. In the present contribution,
we try to extend this successful approach to nonlinear shear
deformations. In contrast to solid state physics,
nonlinear deformations are the rule rather than the
exception in polymer networks and living cells and tissues. 
Most work done so far to understand the mechanical
properties of biopolymer gels 
has focused on crosslinked networks
composed of independent wormlike chains,
disregarding entanglements
\cite{nonlineargels,network_kabla,tissue_WLC_05kuhl,
mackintosh2,frey2,frey3,bendingdominated,mackintosh1,
entropiciscrap,rodney-fivel-dendievel:2005,das-mackintosh-levine:2007,
didonna-lubensky:2005}. 
For realistic fluctuating polymer networks,
the ground state and
linear mechanical properties 
have only very lately been addressed on a molecular basis
\cite{hinsch-wilhelm-frey:07}. 
The nonlinear response was first studied in
the seminal work of Morse extending the Doi-Edwards theory
\cite{tube:morse3}. For filaments with lengths 
much longer than the persistence length,
$L\gg\ell_p$, a softening response with a
weak instability is obtained. The response to
large deformations is dominated by
``hairpins'', filament segments with a large
curvature in the ground state. 
However, the theory provides
neither concrete results nor a physical picture
at the mesh size scale
in the biologically relevant stiff-rod regime, $L\lesssim\ell_p$. 
Here, we aim to account for the full nonlinear elastic
response in a schematic 3D unit-cell of an entangled network
of freely-sliding stiff rods, to obtain
an analytically accessible mean-field description in which the
relevant physical mechanisms can be readily grasped.
In particular, our approach allows for a more straightforward
implementation of finite length effects and clearly
elucidates the origin of the drastic mechanical difference
between entangled polymer networks and enthalpic
fiber scaffolds, which if often underestimated
\cite{doikuzuu,mackintosh2,frey2}.

For the schematic representation of the complicated geometry of
nonlinear deformations we follow the pioneering work by Doi and Kuzuu
\cite{doikuzuu} for enthalpic rod networks. We borrow their simplified
unit-cell approach, including the affinity assumption for the
``background'' of a test chain. This renders the problem analytically
tractable while preserving essential non-affinities caused by the
network geometry on the molecular scale.  The model by Doi and Kuzuu
disregards the non-trivial physics of entanglement, which arises from
the highly correlated thermal fluctuations. To include this crucial
ingredient on an elementary level, we replace their enthalpic rods of
fixed diameter by ``tubes'' (in the sense of the tube model of
semiflexible polymers) of homogeneous but state-dependent width
$d$. Thereby, we project the complicated many-body problem underlying
the phenomenon of entanglement onto a coarse-grained pair problem for
tubes with an effective pair interaction of the form of a Helfrich
repulsion \cite{helfrich}.

Even on this schematic level, the interplay between the free energy
contributions due to bending and tube-confinement turns out to have
interesting consequences for the stress-strain relation, which hardly
would have been anticipated from qualitative arguments.
Due to contact sliding, nonlinear bending does not play a significant role
even for large deformations.
Moreover, thermal undulations largely compensate the collective
stiffening effect predicted by Doi and Kuzuu \cite{doikuzuu}.
Instead, our unit-cell
model exhibits a broad linear elastic regime.  
We conclude that shear stiffening
cannot arise in a solution of freely sliding biopolymers, 
supporting the idea that weakly
adhesive interactions between filaments
are behind the responses observed 
by Semmrich {\it et al} in pure F-actin solutions \cite{christine,nonlinear_semmrich08}.
More interestingly, a
geometric instability is discovered at shear deformations on the order of
one.  The corresponding ``run-away'' effects could be indicative of a
shear-induced structural transition of the polymer solution.  We
speculate that this might then provide an explanation for the
frequently reported collapse of \emph{in-vitro} biopolymer solutions
and networks under shear \cite{christine,nonlinear_semmrich08,bernd06}.  

\section{The model}

\subsection{Overview}

\begin{figure}[p]
\begin{center}
\includegraphics[width=0.4\textwidth]{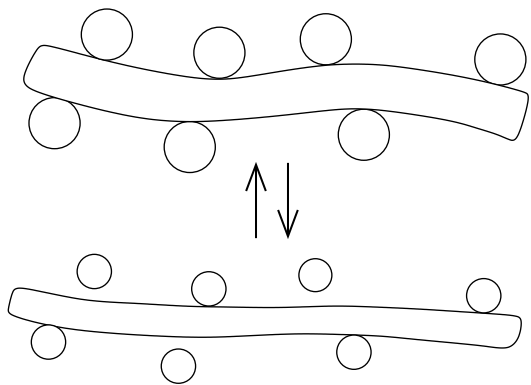}
\caption{ The test tube configuration is a compromise between bending
  and confinement free energy for \emph{fixed} positions of the
  contacts with the surrounding tubes.  }
\label{fig:eq}
\end{center}
\end{figure}

The effective building blocks of an entangled semidilute solution of
semiflexible polymers are thermal tubes \cite{review_kroy}.  Their free
energy can be decomposed into two contributions: the confinement of
long wavelength undulations of the ``enclosed'' polymer, and tube
bending.  Any simplified analytical approach relies on a crucial
assumption on how macroscopic strains couple to the individual
tube. Following Doi and Kuzuu \cite{doikuzuu} but replacing their rods
by tubes, we envision the typical unit-cell as a test tube interacting
via freely slipping entanglements with a background of other,
identically behaving tubes.  The test tube is assumed to be in a local
equilibrium for the given, fixed configuration of the background (Fig.\
\ref{fig:eq}).  Thus the model describes the behaviour on timescales
intermediate between the terminal relaxation time and the typical
slipping time for the contacts. The fact that the slipping time may be
very sensitive to details of the molecular interactions and ambient
conditions \cite{christine} raises interesting prospects for future
developments of the model which we do not pursue further here.



\begin{figure}[p]
\begin{center}
\resizebox{0.7\columnwidth}{!}{
\includegraphics{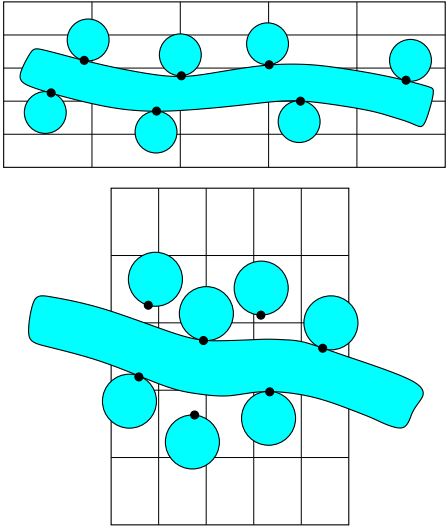}
}
\caption{\label{fig:deform} The background deformation is assumed to
  be affine and volume-preserving. The tube adapts adiabatically to
  its local free energy minimum.  Note that the tube deforms
  non-affinely although the contact points obey the macroscopic
  (affine) deformation.  The number of contacts may also change as the
  background is distorted \cite{doikuzuu}.  }
\end{center}
\end{figure}

The major simplifying assumption of the unit-cell model is that the
background of the test polymer is bound to deform affinely with the
macroscopic deformation given by a deformation tensor ${\bm\Lambda}$.
Due to the discrete nature of the contacts, the test filament deforms
differently from its surroundings.  In this way, nonaffinity at the
entanglement scale is taken into account in an efficient and
transparent approximation (Fig.\ \ref{fig:deform}).  The affine
background has to maintain a constant volume to respect the
(effective) incompressibility of the solvent --- in terms of the
eigenvalues $\lambda_i$ of the deformation tensor ${\bm\Lambda}$,
\[
\lambda_1\,\lambda_2\,\lambda_3 = 1 \;.\] 
As a further, merely technical
simplification \cite{doikuzuu} we assume filaments to lie along the
eigenvectors of the deformation tensor with a periodic arrangement of
contacts with a wavelength given by the mesh size. This simplification
has recently been shown to introduce only very minor errors into the
equilibrium properties of semiflexible polymer solutions
\cite{hinsch-wilhelm-frey:07} and may be expected to remain uncritical
for our purpose. It has the great benefit of allowing for a simple
intuitive explanation of the relevant physics in terms of a few
paradigmatic cell configurations.  Namely, the ensemble of unit-cells
to be considered reduces to the six independent realisations depicted
in Fig.\ \ref{fig:unitcell} (see below).  Considering the full
nonlinear expressions for bending and confinement for these
representative realisations, we can then study large deformations.

\subsection{The Tube}

The free energy
\[
\mathcal{A}= \mathcal{A}^{\rm co} + \mathcal{A}^{\rm be}
\]
of the test tube is split into the contributions from confinement and
bending, $\mathcal{A}^{\rm co}$ and $\mathcal{A}^{\rm be}$,
respectively.  The confinement free energy of a weakly bending rod
constrained by a tube-like harmonic potential of strength $\kappa/(2
L_e^4)$ is $k_B T$ per collision or ``entanglement'' length $L_e$
\cite{odijk}.  The real constraining potential is clearly not uniform,
but rather a discrete set of surrounding tubes.  Nevertheless, since the
main contribution to the confinement energy comes from the constraint
on wavelengths longer than the entanglement length, we may
coarse-grain over the contact points and treat the tube as
homogeneous, with a (state-dependent) half width
\[
\sqrt{\langle \bm{r}_{\!x}^2 (L_e)\rangle} \equiv d/2
\simeq
L_e^{3/2}\ell_p^{-1/2} \;.
\] 
For the remainder, we prefer to work in \emph{natural units} measuring
energies in units of thermal energy $k_B T$ and lengths as multiples
of the persistence length $\ell_p$.  The confinement free energy of a
tube segment of length $\ell$ then takes the form
\[
\mathcal{A}^{\rm co}(d,\,\ell)  \simeq
 \:\ell\,  d^{-2/3}  \;.
\]
And the bending energy of
the space curve $\bm{r}(s)$ of length $\ell$ representing the tube
backbone reads
\[
\mathcal{A}^{\rm be}[\bm{r}(s)] = 
\frac{1}{2} 
\int\limits_0^\ell {\rm d} s \Bigg(\frac{ {\rm d}^2 \bm{r} }{{\rm d}
  s^2 }\Bigg)^2\;.
\]
Taking tube backbones to be {\it elastica} \cite{love}, the bending
energy is automatically at its minimum for given boundary conditions.

\begin{figure}[p]
\begin{center}
\resizebox{0.7\columnwidth}{!}{
\includegraphics{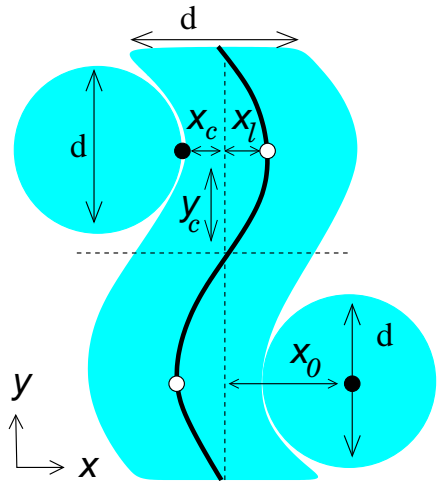}
}
\caption{\label{fig:tube}\small A tube interacting with the
  surrounding tubes.  The tube width is $d$ and its deflection is
  $x_\ell$.  Taking advantage of the simple geometry, the cartesian
  coordinate system is chosen with $x$ along the force $F$ and $y$
  perpendicular to the force along the (average) tube axis.  Two
  alternative versions of the central assertion are considered, namely
  that (1) the contact point (black dot) coordinates $x_c$, $y_c$ or
  (2) the obstacle centers $x_o$, $y_o$ vary affinely with the
  macroscopic strain. }
\end{center}
\end{figure}

We intend to describe networks of strongly entangled filaments for
which both the total contour length and the persistence length are
much longer than the mesh size $\xi$. End-effects at the edges of the
tube can thus be neglected.  Moreover, for our mean-field theory we
may take a periodic arrangement of contacts with a typical distance
$\xi$.  Since the contacts slip freely, the force of magnitude $F$
acting at these contacts must be perpendicular to the test tube.  The
simple geometry naturally suggests the cartesian coordinates
illustrated in Fig.~\ref{fig:tube}.  The coordinate along the force is
$x$, and the coordinate along the (average) tube axis is $y$.  The
point symmetric obstacles have (up to a sign) their center at $x_o$,
$y_o$ and contacts with the test tube at $x_c$, $y_c$. Their relation
to the backbone coordinates $x_\ell,\,y_\ell$ at a ``chemical'' or
arclength distance $\ell$ from the center is read off from
Fig.~\ref{fig:tube},
\begin{eqnarray}
  x_\ell &=& x_o \pm d = x_c \pm d/2 \\
  y_\ell &=& y_0 = y_c \;,
\end{eqnarray}
the sign $\pm$ referring to the sign of $y_c$. Because of the
symmetry, it is sufficient to concentrate onto the positive sector
$y_c>0$ with $\pm \to +$.

The initial position of the contacts has an important influence on the
elastic response of the unit-cell. The mean separation
between contacts along the tube is taken as the entanglement
length, $2 y_c^0=L_e$. 
In the direction perpendicular to the tube
we assume $ x_c^0=d/2$. 
Borrowing
the equilibrium values for entanglement length and tube
diameter from the analysis of the ground state by Hinsch
{\it et al} \cite{hinsch-wilhelm-frey:07},
our ground state conditions are
\begin{subequations}\label{eqs:initial}
\begin{align}
x^0_c &= - \frac{0.32}{ \sqrt{2}}\, \xi^{6/5}  \\  
y^0_c &= + \frac{0.66}{2}\, \xi^{4/5}  \quad. 
\end{align}
\end{subequations}
For typical \emph{in vitro} actin solutions the aspect ratio $y_c^0 /
x_c^0$ has a quite large value of about 7 as a consequence of the
small ratio $\xi/\ell_p\simeq 0.05$. We remark that this
choice of the ground state is only meant as a starting point.
We will eventually explore how much the inital aspect ratio
affects the stress-strain relation.

We now turn to the calculation of the bending energy for arbitrary
states.  Within our mean-field description it suffices to consider the
problem constrained to the plane.  The direction of the force $\bm{F}$
is the natural reference for the tangent angle $\theta$, given by
\[
\bm{F} \cdot \frac{{\rm d}\bm{r}}{{\rm d}s} = F \cos(\theta)\;,
\]
where $F=\vert\bm{F}\vert$ is the magnitude of the external force (in
our natural units).  The equilibrium equation for an Euler-Bernoulli
beam lying in a plane and in absence of body forces is
\begin{equation}\label{eq:pendulo}
\frac{\rm d^2\theta}{{\rm d} s^2} = 
F
\sin \theta
\;,
\end{equation}
the well-known equation of {\it elastica} \cite{love}.  Its first
integral is
\begin{equation}\label{eq:1int}
\frac{1}{2}\,
\Bigg(\frac{\rm d\theta}{{\rm d} s}\Bigg)^{\,2}=
F
\, \Big(\, \cos(\theta_0)-\cos(\theta) \,\Big)
\;.
\end{equation}
So-called inflexional elastica arise in our theory.  Taking an
inflexion point (where ${\rm d}\theta/{\rm d}s=0$) as origin, the
solution to Eq.\ \ref{eq:1int} can be shown \cite{love} to be given by
\begin{subequations}\label{eqs:sol}
\begin{align}
\label{eq:sol1}
\cos(\theta/2) &= k\, \sin(\phi) \\
\label{eq:sol2}
\phi &= {\rm am} \big( s \sqrt{F}
                      + \mathcal{K};\,k \big)   \;, 
\end{align}
\end{subequations}
where ${\rm am}(\:;k)$ is the Jacobi amplitude function with modulus
$k$, and the quarter-period $\mathcal{K}$ is implicitly defined by 
\[ {\rm am}(\mathcal{K};\,k) = \pi/2 \;.\]
The functional form of the solution is given by $k$,
and can be therefore completely defined by
the angle $\theta=\theta_0$ between beam and force at the
inflexion point at the origin, where
\begin{equation}\label{eq:theta0}
s=0\;, \qquad \phi=\pi/2\;, \qquad \cos(\theta_0/2) = k \;,
\end{equation}
holds. That is, the magnitude of the force enters merely as a scale factor
for the arclength coordinate $s$ (in our natural units, $F^{-1/2}$ is
roughly the arc length that would buckle under the load $F$).

\begin{figure}[p]
\resizebox{0.7\columnwidth}{!}{
  \includegraphics*[width=0.25\textwidth]{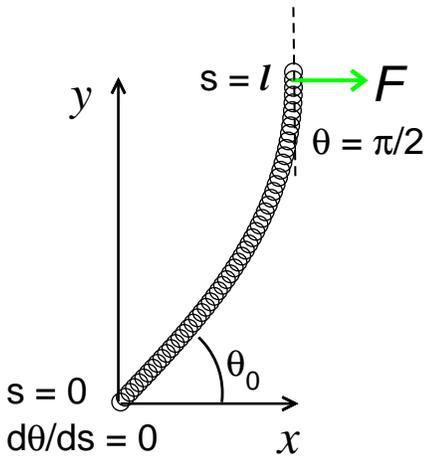}
}
\caption{\label{fig:elastica2} \small Boundary conditions for the tube
  backbone.  At the origin, $s=0$, symmetry requires zero curvature,
  $d\theta/ds=0$.  Since the entanglement at $s=\ell$ slips freely,
  the force $F$ must be perpendicular to the tangent; i.e.,
  $\theta_\ell = \pi/2$.  }
\end{figure}

It will be advantageous to introduce
cartesian coordinates (see Fig.\ \ref{fig:tube}) satisfying
\begin{subequations}\label{eqs:carcoo}
\begin{align}
\frac{{\rm d} x}{{\rm d} s}&=\cos(\theta) \\[1ex]
\frac{{\rm d} y}{{\rm d} s}&=\sin(\theta) \;.
\end{align}
\end{subequations}
The solution can then be written \cite{love} as
\begin{subequations}
 \begin{align}
\label{eq:solx}
 x_s &= 
 s - 
 \frac{2}{\sqrt{
 F} }
 \,\Bigm( \mathcal{E}(\phi\,;k)\: 
 - \: \mathcal{E}(\pi/2;\,k)\Bigm)\\
 \label{eq:soly}
 y_s &= 
 -\frac{2\, k}{ \sqrt{
 F} }
 \:\cos(\phi) \;,
 \end{align}
 \end{subequations}
 where $\mathcal{E}$ is the elliptic function of the second kind.
Setting initial conditions $(\theta,\, {\rm d}\theta/{\rm d}s )$ at a
given point $s$ gives a unique solution for the {\it elastica} shape.
However, the problem is rather one of fulfilling boundary conditions.
As illustrated in Fig. \ref{fig:elastica2}, they are
\begin{subequations}\label{eqs:bc}
\begin{align}
\frac{\rm d\theta}{{\rm d} s} = 0
\quad&\mbox{at}\quad s=0 \label{eq:bc1}\\
\theta = \frac{\pi}{2}
\quad&\mbox{at}\quad s=\ell\;. \label{eq:bc2}
\end{align}
\end{subequations}
The first boundary condition is already implicit in the form of the
solution.

A simple expression for the bending energy of an elastica segment can
be written in cartesian coordinates.  Integrating Eq.\ref{eq:1int}
from $s=0$ to $s=\ell$ gives
\begin{equation}\label{eq:abend}
\mathcal{A}^{\rm be}=F\,\bigg( \ell\, \cos(\theta_0)  - x_\ell \bigg)\;.
\end{equation}
The tube changes its width and distorts its backbone to reach a free
energy minimum for prescribed (affine) coordinates $x_c,\,y_c$ of the
contact point.  (Alternatively, if the coordinates $x_o$, $y_o$ are
taken to obey the affine deformation, they have to be substituted for
$x_c,\,y_c$ in the following discussion.)  The free energy extremum
equation is thus
\begin{equation}\label{eq:eq}
\frac{\partial\mathcal{A}}{\partial{x_\ell}}(x_\ell,x_c,y_c)\Big|_{x_c,\,y_c} = 0
\end{equation}
which gives the equilibrium coordinate $x_\ell^{\rm eq}(x_c,\,y_c)$ of
the tube backbone.  In principle, for the free energy minimization we
thus have to remove $F$, $\ell$ and $\theta_0$ from $\mathcal{A}^{\rm
  be}$ in Eq.~(\ref{eq:abend}) in favour of $x_c$, $y_\ell=y_c$, and
$x_\ell$. This would require a full inversion of the elliptic
functions in Eq.\ \ref{eq:bc2.2}. Therefore, in practice, the
derivative $\partial_{x_\ell}$ has to be rearranged into derivatives
with respect to $F$,
\[
\frac{\partial\mathcal{A}}{\partial x_\ell}^{\!\rm be}\Big|_{y_\ell} =
\frac{\partial\mathcal{A}}{\partial F}^{\!\rm be}\Big|_{y_\ell} \Bigg(
\frac{\partial x_\ell}{\partial F}\Big|_{y_\ell} \Bigg)^{-1} \;,
\] 
where all partial derivatives are taken for constant
$y_\ell$. Accordingly, explicit expressions are only needed for
$x_\ell(F,\, y_\ell)$ and $\mathcal{A}^{\rm be}(F,\, y_\ell)$.

\begin{figure}[p]
\begin{center}
\resizebox{0.75\columnwidth}{!}{
\includegraphics{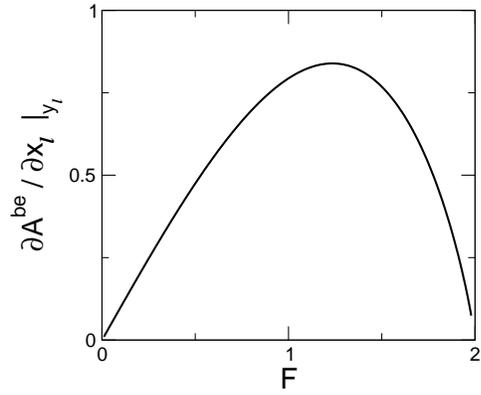}
}
\caption{\small
\label{fig:dABEdx}
Derivative $\partial\mathcal{A}^{\rm be}/\partial x_\ell |_{y_\ell}$
as a function of force $F$ at the contact point (in units $k_b T
\ell_p / y_\ell^2$).  Only in the linear regime they can be
identified.  For large deflections $x_\ell\gg y_\ell$, the force
saturates at $2 k_B T \ell_p / y_\ell^2$ and the bending energy no
longer grows with the backbone coordinate $x_\ell$. }
\end{center}
\end{figure}

The remainder of the section is devoted 
to obtaining expressions in terms of $F,\,y_\ell,\,x_c$
as independent variables. 
Setting $s=\ell$, inverting Eq.\ \ref{eq:soly} and replacing $\phi$ in
Eq.~\ref{eq:solx} gives an expression for $x_\ell$ as a function of
$k,\,F,\,y_\ell$.  Inserting the second boundary condition, Eq.\
\ref{eq:bc2}, into Eqs.\ \ref{eqs:sol} gives
\begin{equation}
\label{eq:bc.2}
\frac{1}{\sqrt{2}} = k \: \sin{ {\rm am} \big(\ell \sqrt{F
  } + \mathcal{K};\,k\big) }\;.
\end{equation}
The (variable) length $\ell$ can be removed by squaring Eqs.\
\ref{eq:soly} and \ref{eq:bc.2} and adding them into
\begin{equation}
\label{eq:bc2.2}
k^2=\frac{1}{2}+ \frac{y_\ell^2\,F}{4}\;. 
\end{equation}
This equation can be used in all elliptic functions to replace the
modulus $k$ by $F$ and $y_\ell$.  It is then straightforward to find
an analytic (albeit cumbersome) expression for $x_\ell(F,\, y_\ell)$.
Similarly, Eqs.\ \ref{eq:bc.2} and \ref{eq:theta0} can also be readily
inverted to obtain expressions for $\ell$ and $\theta_0$.  In this way
we obtain the desired explicit expression for $\mathcal{A}^{\rm
  be}(F,\, y_\ell)$. Figure \ref{fig:dABEdx} shows
$\partial\mathcal{A}^{\rm be}/\partial x_\ell |_{y_\ell}$ as a function of
$F$. The two are equal for small deformations, $x_\ell \ll y_\ell$.
In this linear regime, sliding of the contact point is irrelevant
since the lever arm barely changes, i.e., $\ell \simeq y_\ell$.
However, as the force approaches the order of magnitude of the
spring constant,
$F \sim y_\ell^{-2}$, the contact point
begins to slide noticeably.
As $\ell\rightarrow\infty$,
force and derivative of the bending energy uncouple:
\[ F \rightarrow 2\, y_\ell^{-2} \quad,\quad\quad
\frac{\partial\mathcal{A}}{\partial x_\ell}^{\!\rm be}\Big|_{y_\ell} 
\rightarrow 0 \;.\]
Since the thermodynamic stress is
given by the derivative of the energy,
such nonlinear sliding effects
can asymptotically lead to a peculiar
state where the stress is zero although
the filaments are under lateral tension, $F\neq0$.

A similar rearrangement is necessary for the derivative of the 
confinement energy, since it depends on $\ell$:
\[
\frac{\partial\mathcal{A}}{\partial x_\ell}^{\!\rm co}\Big|_{y_c}
= 
d^{-2/3}\:
\frac{\partial\ell}{\partial F}\Big|_{y_c}
\Bigg(
\frac{\partial x_\ell}{\partial F}\Big|_{y_c}
\Bigg)^{-1} 
+\,
\ell\; \frac{\partial d^{-2/3}}{\partial d/2 \;\;}
\;,
\]
where we have replaced $y_\ell$ by $y_c$.
By adding the derivatives of bending and confinement
energy, we finally obtain an explicit expression for
$\partial\mathcal{A}/\partial x_c$ as a function of $F,\,y_\ell$.
The free energy minimization is done by numerically solving
Eq.\ref{eq:eq} for $F$ for the given, state-dependent coordinates
$y_c$ and $x_c$.


The ultimate goal is of course the stress-strain relation
$\sigma(\gamma)$, with shear strain $\gamma$.
To find an analytic expression for the shear stress
$\sigma$, we first decompose the derivative as
\[\sigma=
\frac{{\rm d} \mathcal{A}}{{\rm d}\gamma}^{\!\rm eq} = 
\frac{\partial\mathcal{A}}{\partial x_c}^{\!\rm eq}\Big|_{y_c}
\:\frac{{\rm d} x_c}{{\rm d}\gamma} + 
\frac{\partial\mathcal{A}}{\partial y_c}^{\!\rm eq}\Big|_{x_c}
\:\frac{{\rm d} y_c}{{\rm d}\gamma}
\;.
\]
The equilibrium free energy can be written in terms of the equilibrium
tube deflection $x_\ell^{\rm eq}(x_c,\,y_c)$ as
\begin{equation}
\label{eq:eq_free}
\mathcal{A}^{\rm eq}(x_c,\,y_c) = 
\mathcal{A}
(x_\ell^{\rm eq}
{\!}_{(x_c,\,y_c)}   
,\,
x_c   ,\,
y_c    )
\;,
\end{equation}
a function of the macroscopic deformation. Taking the derivative with
respect to the contact position $x_c$ gives
\[
\frac{\partial\mathcal{A}}{\partial x_c}^{\!\rm eq}\Big|_{y_c}
=
\frac{\partial\mathcal{A}}{\partial x_\ell}\Big|_{x_c,\,y_c}
\:
\frac{\partial x_\ell}{\partial x_c}^{\!\rm eq}\Big|_{y_c}
+
\frac{\partial\mathcal{A}}{\partial x_c}\Big|_{y,\,x_\ell=x_\ell^{\rm eq}}
\;.
\]
Since the energy is at its minimum for constant contact coordinates
(Eq.\ \ref{eq:eq}), the first term on the rhs is equal to zero and
therefore
\[
\frac{\partial\mathcal{A}}{\partial x_c}^{\!\rm eq}\Big|_{y_c}
=
\frac{\partial\mathcal{A}}{\partial x_c}\Big|_{y_c,\,x_\ell=x_\ell^{\rm eq}}
=
- \frac{\partial\mathcal{A}}{\partial d/2}^{\!\!\!\!\rm co}\;,
\]
where the last equality holds since $\mathcal{A}^{\rm be}$
does not depend on $x_c$.
One similarly finds
\[
\frac{\partial\mathcal{A}}{\partial y_c}^{\!\rm eq}\Big|_{x_c}
\!=\:
\frac{\partial\mathcal{A}}{\partial y_c}^{\!\rm be}\Big|_{x_\ell}
+
d^{-2/3}  \frac{\partial\ell}{\partial y_c}\Bigg|_{x_\ell}
\;.
\]
We finally obtain
\begin{multline}
\label{eq:stress}
\sigma
\,=\,
- \frac{\partial\mathcal{A}}{\partial d/2}^{\!\!\!\!\rm co}
\; \frac{{\rm d} x_c}{{\rm d}\gamma}
\;+ \\
\Bigg(\,
\frac{\partial\mathcal{A}}{\partial y_c}^{\!\rm be}\Big|_{x_\ell}
+
d^{-2/3}  \frac{\partial\ell}{\partial y_c}\Big|_{x_\ell}
\Bigg)\,
\!\frac{{\rm d} y_c}{{\rm d}\gamma}
\;.
\end{multline}
We will later use this equation to gain physical insight into the
mechanical stability of the solution.  Note that derivatives with
respect to $y_c$ for constant $x_\ell$ require again a rearrangement,
since we do not have explicit expressions with $x_\ell$ as independent
variable.  For this we need
\[ \frac{\partial\mathcal{A}}{\partial y_c} ^{\!\rm be} \Big|_{x_\ell} 
=\:
\frac{\partial\mathcal{A}}{\partial y_c} ^{\!\rm be} \Big|_{F}
- \:
\frac{\partial\mathcal{A}}{\partial F} ^{\!\rm be} \Big|_{y_c}
\:\frac{\partial x_\ell}{\partial y_c} \Big|_{F}
\: \left( \frac{\partial x_\ell}{\partial F} \Big|_{y_c} \right)^{-1}
\]
\[
\frac{\partial\ell}{\partial y_c}  \Big|_{x_\ell} 
=\:
\frac{\partial\ell}{\partial y_c} \Big|_{F}
- \:
\frac{\partial\ell}{\partial F}  \Big|_{y_c}
\:\frac{\partial x_\ell}{\partial y_c} \Big|_{F}
\: \left( \frac{\partial x_\ell}{\partial F} \Big|_{y_c} \right)^{-1} 
\:.
\]

\subsection{The Unit-Cell}

We have defined the $x$, $y$, $z$ coordinates as fixed to an arbitrary
set of tubes.  The test tube lies along $y$ and interacts with
background tubes which lie along $z$ and move in the $x$--$y$ plane.
For an isotropic network, the unit-cell should be modeled as a
superposition of test- and background tubes lying along all possible
orientations, integrating the triad $x$--$y$--$z$ over the whole 3D
triad-space. However, this would complicate 
the problem enormously, since triads lying oblique to the
main axis of the deformation would distort.
To simplify matters
we follow Doi and Kuzuu \cite{doikuzuu} and
distribute test- and background tubes only along the main stretch
directions (the eigenvectors of the right stretch tensor \cite{topics}).
Since then tubes 
remain perpendicular to each other,
the deformation can be incorporated in the
analytical treatment as a translation
of the contact point.
The total free energy of the unit-cell is then a
sum of 6 terms,
\[
\mathcal{A}=
 \mathcal{A}_{12}
+ \mathcal{A}_{13}
+ \mathcal{A}_{21}
+ \mathcal{A}_{23}
+ \mathcal{A}_{31}
+ \mathcal{A}_{32}
 \quad.
\]
We denote configurations by the coordinates of the deformation plane;
for a particular tube triad, the free energy is given by
$\mathcal{A}_{xy}$.  \footnote[1]{Here we deviate from the notation used
in Ref.\ \cite{doikuzuu}}
 Figure \ref{fig:unitcell} depicts all possible
arrangements.


\subsection{The deformation}

We assume the contact points to deform affinely,
going from their initial positions $\bm{{\rm R}}_c$
to final positions $\bm{{\rm r}}_c$.
The assumption of affinity means that
the deformation gradient 
$\bm{\Lambda}=\partial\bm{{\rm r}}_c/\partial\bm{{\rm R}}_c$,
is spatially constant, and thus
\[
\bm{{\rm r}}_c = \bm{\Lambda\,{\rm R_c}}
\]
holds for all contact points. 
With our assumption that the tubes lie along
the stretch directions,
this statement can be written 
in terms of the stretch factors as
\[\begin{split}
x_c &= \lambda_x\, x_c^0 \\
y_c &= \lambda_y\, y_c^0  \;.
\end{split}\]
Solving Equation \ref{eq:eq} we obtain the equilibrium free energy
$\mathcal{A}^{\rm eq}$ as a function of the deformation.  The (Cauchy)
stress tensor $\bm\sigma$ can then be calculated as
the derivative of the equilibrium free energy per unit volume,
\[
\bm\sigma=
\det\bm{\Lambda}^{-1}
\frac{\partial\mathcal{A}^{\rm eq} / V_e}
{\partial\bm{\Lambda}}
\bm{\Lambda}^T  \;,
\]
where the volume of the unit-cell is taken as
\[
V_e =  \xi^2 L_e \simeq  x_c^0 \,{y_c^0\,}^2 \;.
\]
For concreteness and
comparison with experiments we will restrict ourselves to volume
preserving shear deformations in the plane 1--2, with positive stretch
along 1 and contraction along 2. In terms of the eigenvalues of the
strain tensor, $\lambda_1\rightarrow\infty$, $\lambda_2\rightarrow 0$,
and $\lambda_3=1$ when $\gamma\rightarrow\infty$.  The shear stress is
given by
\begin{equation}
\label{eq:stress1}
\sigma=
\frac{\partial\mathcal{A}^{\rm eq} / V_e}
{\partial\gamma} \;.
\end{equation}

Since the ground state is assumed to be the same for all orientations
--- as expected for an isotropic network --- the sum of all stresses
must give $\sigma=0$ for $\gamma=0$ for the unit-cell.  For shear in
the 1--2 plane, it is readily seen that two configurations cancel each
other's stress if they are related by an interchange of the directions
$1$, $2$.

\subsection{Doi-Kuzuu effect}

By ``Doi-Kuzuu effect'' we mean the change in the number of contacts
between filaments due to the deformation \cite{doikuzuu}.  
A parallelepiped with sides $2d/\lambda_x$, $L/\lambda_y$,
$L/\lambda_z$, 
assumed to lie along the 
principal stretches,
distorts affinely into a parallelepiped with sides $2d$, $L$, $L$.  
By construction, 
the number of filaments lying along $z$ which come into
contact with the test rod can be readily shown to be \cite{doikuzuu}
\[
\Delta n =
\frac{c}{3}\frac{L^2}{\lambda_y\lambda_z} \left(\frac{2d}{\lambda_x} -2d \right)
= \frac{2}{3}\,cdL^2(1-\lambda_x) .
\]
Thus, as the solution is distorted at constant volume, the number of
contacts between rods increases, reducing the lever arm
for bending. Assuming stress to be of purely enthalpic nature
leads to a pronounced stiffening response \cite{doikuzuu}. 
We remark that the material becomes stiffer 
without the rods ever leaving the linear
bending regime, a pure network effect. 

Such a general stiffening mechanism is 
certainly worth considering and can be readily
incorporated in our theory. The inverse of the
number of contacts per unit length, $L/n$, 
amounts to the distance $y_c$ between contacts. 
For our purposes, however, the stiff rods 
have to be replaced by thermal tubes
with a variable diameter. For this we recast
the original argument in differential form. 
 Replacing
$cL/3$ by $1/\xi^2$, we get
\[
\delta y_c= \frac{2\, y_c^2 \,d}{\xi^2} \: \delta\lambda_x \;.
\]
An extra term is needed to address 
tube thickening in absence of shear, 
which increases
the number of contacts according to
\[
\delta y_c= - \frac{2\, y_c^2}{\xi^2} \: \delta d \;.
\] 
The two terms influence the evolution
of the distance between contacts, $y_c$.
Within our mean-field approach it is natural
to add them to the affine macroscopic deformation:
\begin{equation}
\label{eq:DK}
\delta y_c = 
y_c^0 \:\delta\lambda_y +
\frac{2 \,y_c^2 \,d }{\xi^2} \:\delta\lambda_x -
\frac{2\, y_c^2}{\xi^2} \: \delta d \;.
\end{equation}

\section{Results}

\subsection{Softening and instability}

Figure \ref{fig:xi} shows the response
of the unit-cell for mesh sizes $\xi/\ell_p=0.01$ and $0.1$. 
The stress-strain relation shows a broad linear
regime which softens
and becomes unstable beyond 100\% shear strain.
Though the theory employs the full nonlinear
equations for bending, stiffening is not observed,
suggesting that the deflection of the test tube 
remains small throughout. Indeed, linearizing
$\mathcal{A}^{\rm be}$ in $x_\ell$ around zero
has only a minor effect on the stress-strain relation,
as shown in Fig.\ \ref{fig:xi}. 
A consequence of the small deflection is that
changing the mesh size $\xi$ essentially amounts
to a scaling of the stress axis, as can be
seen in Fig.\ \ref{fig:xi}. For linearized
bending the equilibrium deflection scales like
$x_\ell^{\rm eq} \sim \xi^{4/5}$, which renders both
confinement and (linearized) bending energy
independent of the mesh size
(both are $\sim k_BT$).
Therefore the stress goes like
\[ \sigma = \frac{{\rm d}\mathcal{A}}{{\rm d}\gamma}^{\rm eq,\,lin}
\!\!\! \!\!\! \!\!\! \!\!\!(\gamma)
\;\Big/\; V_e(\xi) 
 \sim  \xi^{-14/5} 
 \;,\]
which boils down to the well known prediction
of the tube model for the concentration dependence of the shear modulus,
$G_0\sim c^{\,1.4}$. 
Deviations from this simple mesh-size scaling
are a measure of nonlinear bending.

A more important parameter 
is the initial aspect ratio, $y_c^0/x_c^0$; as shown in Fig.\
\ref{fig:rA}, it decides on the stability of the response.
A unit-cell with an aspect ratio larger than $\sim 10$
is already unstable in the ground state. 
Smaller aspect ratios push the instability away,
but without a significant increase in the modulus.

To better understand the mechanical response of the unit-cell in terms
of the microscopic deformations, it is helpful to consider the
individual contributions of each $xy$ orientation 
(Fig.\ \ref{fig:richtig}).
In general, none of the individual configurations displays stiffening
behaviour. To a varying extent the instability shows up in all
contributions.  It appears at its strongest in the 2--1-term,
corresponding to a test tube along the stretch direction and
background tubes along the third direction, moving in the plane 1--2
(see Fig.\ \ref{fig:unitcell}).
Since the transverse stiffness of a beam of length $L$ decreases
proportional to $L^{-3}$ with increasing length, an increase in the
mean distance between adjacent contacts $y_c$ weakens the resistance
of the test tube, so that its backbone yields to the pressure exerted
by the surrounding tubes, which try to expand in order to minimize
their confinement free energy.  As the deflection of the test tube
increases, the negative contribution to the stress from
\[
\frac{\partial\mathcal{A}}{\partial y_c}^{\!\rm be}\Big|_{x_\ell}
\]
will eventually dominate Eq.\ \ref{eq:stress}.  The instability is an
inescapable consequence of our \emph{Ansatz} for the background
deformation and the freely sliding contacts.  Figure
\ref{fig:intuitivo} tries to convey an intuitive feeling for it.

One may wonder whether this instability is an unavoidable
feature of stiff polymer solutions or whether higher order
correlations neglected in our simple unit-cell approach will
eventually stop a real solution from yielding. After all, a living
cell should be able to withstand large stretch without
collapsing. In the remainder of this section we will digress on
two mechanisms which provide stability against 
large deformations: the Doi-Kuzuu effect \cite{doikuzuu}
and hairpins \cite{tube:morse3}.


\subsection{Thermal Doi-Kuzuu effect}

The common trend in the results shown so far
is a softening response. This seems at odds
with the experimentally observed stiffening responses 
in F-actin solutions \cite{christine,nonlinear_semmrich08}. 
The question arises whether collective effects
may be behind them, as in the Doi-Kuzuu effect,
which predicts a power-law stiffening response for
a solution of rigid rods. 
As shown in Fig.\ \ref{fig:DK},
replacing the rods by
tubes with a variable diameter has a dramatic
consequence: the stiffening response becomes
a softening one. The tug-of-war between Doi-Kuzuu
reinforcement via increase in entanglement density
and the thermal sliding instability leads
to a neutral result where both largely compensate each other.
Therefore, though unable to provide stiffening,
the Doi-Kuzuu effect renders the unit-cell stable.


\subsection{Hairpins are stable}



Here we point out a simple mechanism to stabilize the
network even on the level of the single unit-cell.  Namely, 
as depicted in Fig.\ \ref{fig:misfit}:
if the contact $x_c$ lies ``on the other side'', $x_c > 0$, such that the
test tube is more strongly bent in the ground state.
This boils down to a microscopic realization of the
``hairpins'' first discussed by Morse \cite{tube:morse3}. 
Though at first sight innocuous, the change in topology
has dramatic consequences for the stress-strain relation.
As shown in Fig.\ \ref{fig:misfit-response}, the response
becomes a weak stiffening one and the sliding instability
vanishes. This weak stiffening is not
a consequence of nonlinear bending, 
as can be seen by the response 
of the linearized theory. Stiffening arises
as the distance between contacts and hence the lever
arm $y_c$ decreases, an effect which dominates for the
configuration 1--2 (see Fig. \ref{fig:misfit-response}, inset).
The nonlinear theory actually tames the response
into softening at very large shear. This
is a consequence of entering a purely sliding regime
as in Fig.\ \ref{fig:dABEdx}.
Thus, though the linearized theory 
suffices for the typical unit-cell,
it grossly overestimates stiffening
in the hairpin configuration.

\section{Discussion}

In the present work we have developed
a simple, analytically tractable mean-field 
description of solutions of stiff biopolymers,
describing an adiabatic
evolution at timescales longer than
typical times for crosslink slippage. 
The theory makes use of the full nonlinear
bending equations and is hence 
suitable to address large deformations.
Surprisingly, in spite of all
stiffening mechanisms considered, 
the theory invariably predicts a broad linear 
regime which softens and eventually becomes
unstable at large shear strain.
Nonlinear bending, which
leads to strong stiffening responses
in the fixed-contact scenario \cite{mipaper},
becomes largely irrelevant when contacts slide. 
The hairpin configuration, though stable,
gives a very weak stiffening response which
does not resemble the experimental data,
where the linear regime does not extend beyond 5\% strain
\cite{christine,nonlinear_semmrich08}. 
Finally, our version for semiflexible polymers of the geometric
Doi-Kuzuu stiffening mechanism \cite{doikuzuu}
is largely cancelled out by thermal fluctuations. 
Though it does provide stability, 
the response remains a softening one. 
We conclude that stiffening in biopolymer solutions
necessarily requires coupling to the longitudinal
stretching and compression modes of the individual filaments.
Our conclusion supports the previous interpretation 
of stiffening in terms of the glassy wormlike chain \cite{christine}.
Understood as a consequence of friction,
the fact that stiffening requires fast
rates and low temperatures and leads to large dissipation
makes indeed sense.

Having gained this insight, 
we can outline a
full theory of nonlinear viscoelasticity
of biopolymer gels.
At fast rates, effectively 
the solution behaves like a network;
here one may expect theories developed for
crosslinked networks to hold. 
Importantly, the elastic strain
will be limited to values below
$\sim5\%$ strain, since at larger strains
crosslinked networks stiffen dramatically.
The large forces attained by the stiffening filaments
will break bonds faster
and render the deformation inelastic.
Thus, for finite rates we expect
the nonlinear response
to go in hand with huge structural dissipation.
As the rate is lowered to timescales 
intermediate between crosslink slippage
and terminal relaxation, the present theory should hold,
predicting an
almost reversible response, 
with a broad linear regime.
This picture is indeed corroborated
by experiments. 
Both for living cells \cite{monolayer,fernandez2,modeltissue,linear_saif}
and in-vitro F-actin solutions \cite{christine,nonlinear_semmrich08},
stress-strain relations 
become increasingly linear over large deformations
as the strain rate is lowered,
and the reversibility of the response increases
as dissipation goes down.
Thus, there seems to be hope of a unified description
of this plethora of mechanical responses
in terms of networks of semiflexible
filaments with frictional interactions.

As the deformation spreads the contacts
along the main stretch axis,
their lever arm increases and the test tube
no longer can sustain
the thermal fluctuations. This leads to
a sliding instability and presumably to shear
banding at a critical strain of 100\%.
This picture provides a fresh, novel look at the mysterious
``collapse'' of biopolymer networks and solutions.
Pure F-actin solutions undergo irreversible 
weakening at $\sim150\%$
\cite{christine,nonlinear_semmrich08}.
The experimental fact that this collapse 
takes place at a well defined
strain, largely independent of rate and stress, 
makes the case for a purely geometrical instability, 
unrelated to breakage of filaments, possibly
resembling collective rearrangements
in oil tubes \cite{collapse_teresa}.
Though the full nonlinear theory including
the coupling to the longitudinal polymer modes
is still under development \cite{hybrid_kroy08},
the reasoning outlined above supports the sliding stability
as a qualitative explanation for the collapse.
Indeed, since the sliding instability
takes place at strains $\sim 100\%$ in our theory, 
and the backbone strain 
cannot be much larger than 5\%,
one would expect the critical collapse strain
to be largely unaffected by the
degree of stiffening, i.e., by the strain rate
-- as experimentally observed.

Hairpin-entanglements are interesting, 
as they are stable against large deformations. In fact,
these hairpins are always present with a certain small
probability on the scale of the unit-cell, and with higher probability
on larger scales. 
Closely related to the hairpins described by Morse 
in long filaments \cite{tube:morse3},
they have a drastic effect onto the mechanical
response, because their deformation and free energy invariably
increase upon affine background deformations. 
By addressing hairpins and normal entanglements separately, their
opposite responses at large strains stand out.
The instability of the unit-cell against shear may be turned into a
weak stiffening response upon changing 
the entanglement topology to a hairpin.
One may therefore wonder whether hairpins are
behind the fact that the conspicuous in-vitro collapse is not observed
when straining living cells by over 100\%
 \cite{fernandez2,monolayer}. 
Since large deformations of this order
are to be expected under physiological conditions 
(as cells crawl and divide), it seems
desirable to avoid a collapse of the main load-bearing
structure in the cell. 
Though typical actin filament sizes are
probably too short to provide strongly bent tubes, 
one may create hairpins in alternative ways.
To achieve the right ``inside-out'' topology,
it would suffice to crosslink filaments at a large angle
and let them polymerize, thereby effectively 
generating a large bent in the ground state.
Such a strategy would remind of the Arp 2/3 complex, an
ubiquitous component of the actin cortex.
As a more generic possible mechanism
one can imagine a pre-bending of filaments in
the ground state enforced by mutual adhesive interactions
or simply by their embedding into a disordered environment.
This would provide an intimate link between the question
of mechanical stability of living cells and tissues
and a fundamental polymer physics problem.

%

\vspace{1cm}

We are very grateful to Andreas R. Bausch
for his generous support.
Insightful discussions with C. Semmrich and J. Glaser
are also acknowledged.

\clearpage

\bibliography{/Users/pablo/biblio/cellpulling}

\clearpage
\begin{list}{}{\leftmargin 2cm \labelwidth 1.5cm \labelsep 0.5cm}

\item[\bf Fig. 1] The test tube configuration is a compromise between bending
  and confinement free energy for \emph{fixed} positions of the
  contacts with the surrounding tubes. 
\item[\bf Fig. 2] The background deformation is assumed to
  be affine and volume-preserving. The tube adapts adiabatically to
  its local free energy minimum.  Note that the tube deforms
  non-affinely although the contact points obey the macroscopic
  (affine) deformation.  The number of contacts may also change as the
  background is distorted \cite{doikuzuu}.  
\item[\bf Fig. 3] A tube interacting with the
  surrounding tubes.  The tube width is $d$ and its deflection is
  $x_\ell$.  Taking advantage of the simple geometry, the cartesian
  coordinate system is chosen with $x$ along the force $F$ and $y$
  perpendicular to the force along the (average) tube axis.  Two
  alternative versions of the central assertion are considered, namely
  that (1) the contact point (black dot) coordinates $x_c$, $y_c$ or
  (2) the obstacle centers $x_o$, $y_o$ vary affinely with the
  macroscopic strain. 
\item[\bf Fig. 4] Boundary conditions for the tube
  backbone.  At the origin, $s=0$, symmetry requires zero curvature,
  $d\theta/ds=0$.  Since the entanglement at $s=\ell$ slips freely,
  the force $F$ must be perpendicular to the tangent; i.e.,
  $\theta_\ell = \pi/2$. 
\item[\bf Fig. 5] Derivative $\partial\mathcal{A}^{\rm be}/\partial x_\ell |_{y_\ell}$
as a function of force $F$ at the contact point (in units $k_b T
\ell_p / y_\ell^2$).  Only in the linear regime they can be
identified.  For large deflections $x_\ell\gg y_\ell$, the force
saturates at $2 k_B T \ell_p / y_\ell^2$ and the bending energy no
longer grows with the backbone coordinate $x_\ell$. 
\item[\bf Fig. 6] Unit-cell for shear in the 1--2 plane.  The main stretch directions
1--2--3 rotate with the current configuration.  The unit-cell is
modeled as a superposition of test tubes lying along the main
stretches.  Each configuration is named according to the $x$--$y$
coordinates.  The arrows indicate the displacement of the contacts
upon an increase of the affine strain $\gamma$.  
\item[\bf Fig. 7] Normalized shear stress-shear strain relation
for mesh sizes $\xi/\ell_p=0.01$ (solid lines) and $0.1$
(dashed lines). The mesh size barely makes a difference,
a consequence of the small filament deflection. 
The deformation of the background has been
considered in two ways.
Gray lines: middle points deform affinely,
$x_0 = \lambda_x x_0^0$.
Black lines: contact points deform affinely,
$x_c = \lambda_x x_c^0$.
Dotted line: linearized bending energy.
For the linearized theory, changing the mesh size
merely scales the stress axis.
\item[\bf Fig. 8] Aspect ratio dependence.  Shear stress-shear strain
  relation for mesh size $\xi = 0.02\,\ell_p$.  The initial aspect ratio $y_c^0
  / x_c^0$ increases along the arrow, taking the values 3.5, 4.9, 7,
  10, 14, with $y_c^0 / x_c^0\approx 7$ corresponding to the tube
  model, Eq.\ \ref{eqs:initial}. 
\item[\bf Fig. 9] Shear stress versus shear strain for all 6 possible
  orientations of the test and background tubes, for mesh
  size $\xi=0.02\,\ell_p$. The curve passing
  through zero stress is the sum of all others, representing the
  response of the unit-cell.  The instability is essentially due to
  the term 21. 
\item[\bf Fig. 10] Cartoon illustrating the instability by an intuitive analogy.  Due to
incompressibility of the shaded area, the contact point (represented
by a filled circle) can only move
along the dashed line.  The springs (representing confinement free
energy) have an infinite rest length and slide along clamped beams
(representing tube bending). 
Movement of the contact point increases (lowers) the compression on
the spring opposing (favouring) the movement. 
The arrows correspond to
the total force acting on the contact point. In the ground state the lever arm is
short and the beams barely bend -- the system is linearly stable.
 As the network is
sheared further the increase in lever arm softens the beam, whose
bending gives rise to a lateral force pushing out the
  contact. The system becomes unstable.

\item[\bf Fig. 11] Doi-Kuzuu effect.
Shear stress-shear strain relation for mesh size $\xi = 0.02\,\ell_p$,
with and without DK effect.
The stiffening response reported in Ref.\ \cite{doikuzuu}
is ``ironed out'' by the thermal fluctuations.

\item[\bf Fig. 12] A hairpin. Setting $x_c^0 > 0$, while
  still complying with the requirement $\lvert x_c^0 \rvert \lvert
  y_c^0\rvert\simeq\xi^2$, leads to a very different response. Since
  now the tubes are ``trapped'', the entangled solution is stable
  against large deformations.  
\item[\bf Fig. 13] Shear stress versus shear strain
for a hairpin configuration where $x_c>0$, as in Fig.\ \ref{fig:misfit}.
Solid line: full nonlinear theory. Dashed line: linearized theory.
The instability vanishes for the hairpin configuration. Note that
the weak stiffening response is not due to nonlinear bending,
since it is present also in the linearized theory; it is rather
a consequence of the decrease of the distance between contacts.
{\bf Inset:} 
 Shear stress versus shear strain for all 6 possible
  orientations of the test and background tubes. 

\end{list}

\clearpage

\clearpage

\clearpage

\clearpage

\clearpage

\clearpage

\begin{figure}[p]
\begin{center}
\resizebox{0.9\columnwidth}{!}{
\includegraphics{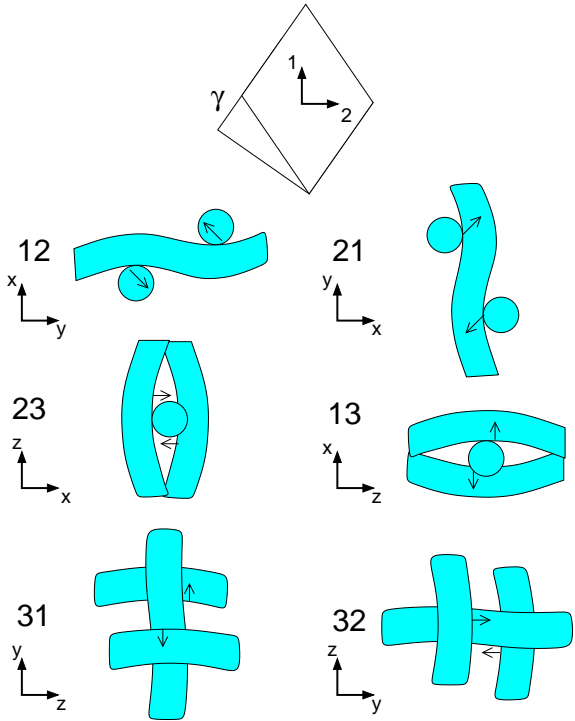}
}
\caption{\small
\label{fig:unitcell}
Unit-cell for shear in the 1--2 plane.  The main stretch directions
1--2--3 rotate with the current configuration.  The unit-cell is
modeled as a superposition of test tubes lying along the main
stretches.  Each configuration is named according to the $x$--$y$
coordinates.  The arrows indicate the displacement of the contacts
upon an increase of the affine strain $\gamma$.  }
\end{center}
\end{figure}

\clearpage

\begin{figure}[p]
\begin{center}
\resizebox{0.95\columnwidth}{!}{
\includegraphics{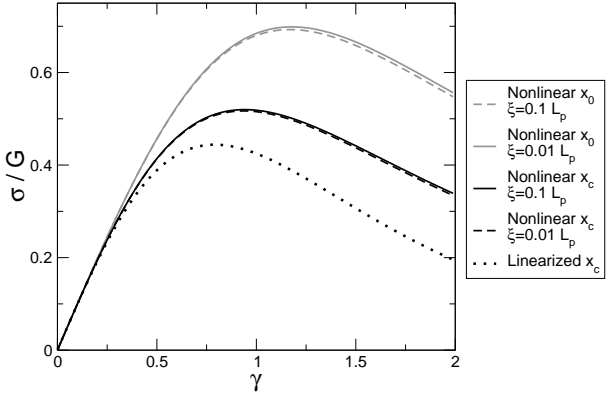}
}\caption{\label{fig:xi} Normalized shear stress-shear strain relation
for mesh sizes $\xi/\ell_p=0.01$ (solid lines) and $0.1$
(dashed lines). The mesh size barely makes a difference,
a consequence of the small filament deflection. 
The deformation of the background has been
considered in two ways.
Gray lines: middle points deform affinely,
$x_0 = \lambda_x x_0^0$.
Black lines: contact points deform affinely,
$x_c = \lambda_x x_c^0$.
Dotted line: linearized bending energy.
For the linearized theory, changing the mesh size
merely scales the stress axis.}
\end{center}
\end{figure}

\clearpage

\begin{figure}[p]
\begin{center}
\resizebox{0.8\columnwidth}{!}{
\includegraphics{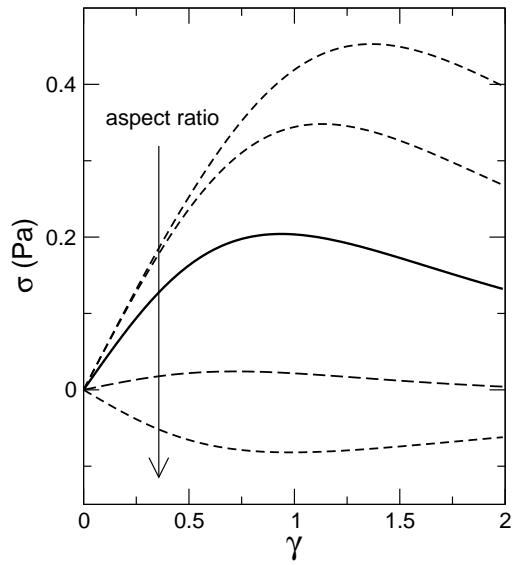}
}
\caption{\label{fig:rA} Aspect ratio dependence.  Shear stress-shear strain
  relation for mesh size $\xi = 0.02\,\ell_p$.  The initial aspect ratio $y_c^0
  / x_c^0$ increases along the arrow, taking the values 3.5, 4.9, 7,
  10, 14, with $y_c^0 / x_c^0\approx 7$ corresponding to the tube
  model, Eq.\ \ref{eqs:initial}.  }
\end{center}
\end{figure}

\clearpage

\begin{figure}[p]
\begin{center}
\resizebox{0.75\columnwidth}{!}{
\includegraphics{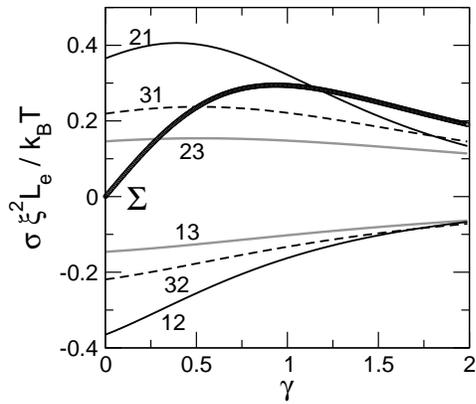}
}
\caption{\label{fig:richtig} Shear stress versus shear strain for all 6 possible
  orientations of the test and background tubes, for mesh
  size $\xi=0.02\,\ell_p$. The curve passing
  through zero stress is the sum of all others, representing the
  response of the unit-cell.  The instability is essentially due to
  the term 21. }
\end{center}
\end{figure}

\clearpage

\begin{figure}[p]
\begin{center}
\resizebox{0.9\columnwidth}{!}{
	\includegraphics[width=0.8\textwidth]{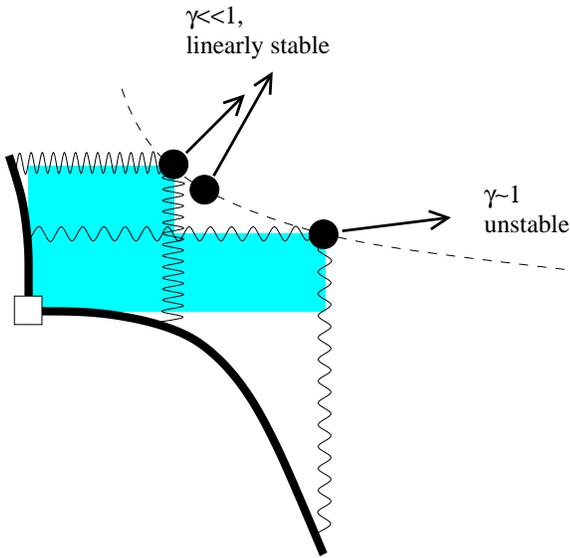}} 
\caption{
\label{fig:intuitivo}
Cartoon illustrating the instability by an intuitive analogy.  Due to
incompressibility of the shaded area, the contact point (represented
by a filled circle) can only move
along the dashed line.  The springs (representing confinement free
energy) have an infinite rest length and slide along clamped beams
(representing tube bending). 
Movement of the contact point increases (lowers) the compression on
the spring opposing (favouring) the movement. 
The arrows correspond to
the total force acting on the contact point. In the ground state the lever arm is
short and the beams barely bend -- the system is linearly stable.
 As the network is
sheared further the increase in lever arm softens the beam, whose
bending gives rise to a lateral force pushing out the
  contact. The system becomes unstable.}
\end{center}
\end{figure}

\clearpage

\begin{figure}[p]
\begin{center}
\resizebox{0.8\columnwidth}{!}{
\includegraphics{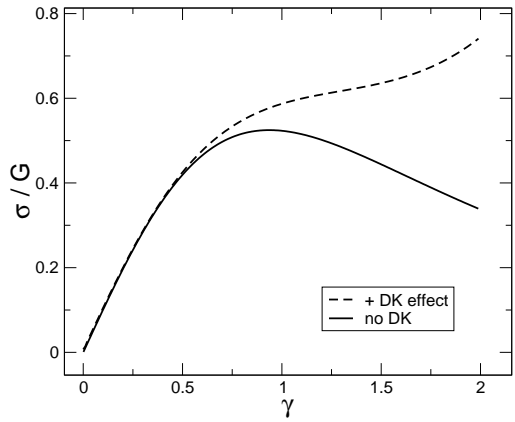}
}
\caption{\label{fig:DK}
Doi-Kuzuu effect.
Shear stress-shear strain relation for mesh size $\xi = 0.02\,\ell_p$,
with and without DK effect.
The stiffening response reported in Ref.\ \cite{doikuzuu}
is ``ironed out'' by the thermal fluctuations.
}
\end{center}
\end{figure}

\clearpage

\begin{figure}[p]
\resizebox{0.9\columnwidth}{!}{
\includegraphics*[width=0.5\textwidth]{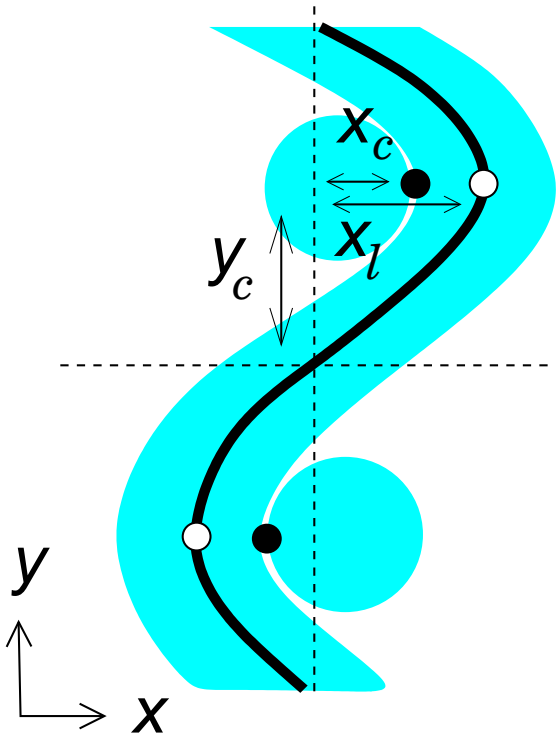}
}
\caption{\label{fig:misfit} A ``hairpin''. Setting $x_c^0 > 0$, while
  still complying with the requirement $\lvert x_c^0 \rvert \lvert
  y_c^0\rvert\simeq\xi^2$, leads to a very different response. Since
  now the tubes are ``trapped'', the entangled solution is stable
  against large deformations.  }
\end{figure}

\clearpage

\begin{figure}[p]
\begin{center}
\resizebox{0.9\columnwidth}{!}{
\includegraphics{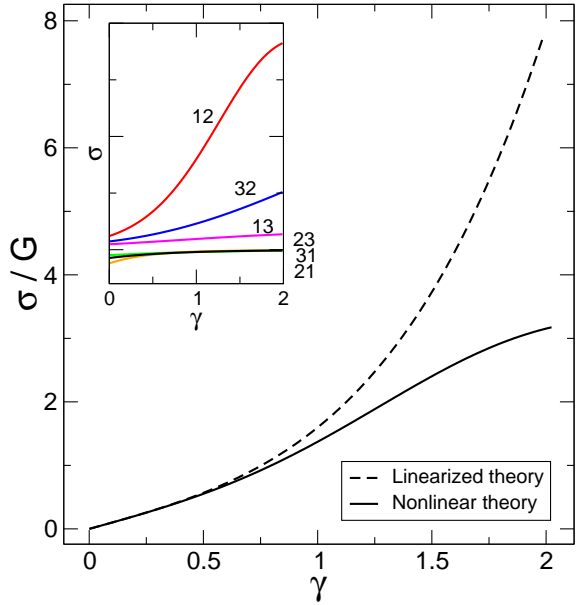}
}
\caption{\label{fig:misfit-response} Shear stress versus shear strain
for a ``hairpin'' configuration where $x_c>0$, as in Fig.\ \ref{fig:misfit}.
Solid line: full nonlinear theory. Dashed line: linearized theory.
The instability vanishes for the hairpin configuration. Note that
the weak stiffening response is not due to nonlinear bending,
since it is present also in the linearized theory; it is rather
a consequence of the decrease of the distance between contacts.
{\bf Inset:} 
 Shear stress versus shear strain for all 6 possible
  orientations of the test and background tubes. }
\end{center}
\end{figure}

\end{document}